\documentclass[showpacs,aps,prl,twocolumn,superscriptaddress]{revtex4-2}

\usepackage{graphicx}
\usepackage{hyperref}
\usepackage{epstopdf}
\usepackage{dcolumn}
\usepackage{bm}
\usepackage{amsmath}
\usepackage{amssymb}
\usepackage[dvipsnames]{xcolor}
\usepackage{color} 
\usepackage{soul}
\usepackage{xspace}

\usepackage{siunitx}

\usepackage{float}

\renewcommand{\vec}[1]{\boldsymbol{#1}}

\newcommand{\ve}[1]{\boldsymbol{#1}}

\setcitestyle{square,numbers}
\usepackage{here}
\usepackage{cancel} 
\usepackage{ulem}

\begin{document}

\title{Scale-invariant magnetic anisotropy in  $\alpha$-RuCl$_3$: A quantum Monte Carlo study}

\author{Toshihiro Sato}
\affiliation{\mbox{Institute for Theoretical Solid State Physics, IFW Dresden, 01069 Dresden, Germany}}
\affiliation{\mbox{W\"urzburg-Dresden Cluster of Excellence ct.qmat, Germany}}
\affiliation{\mbox{Institut f\"ur Theoretische Physik und Astrophysik, Universit\"at W\"urzburg, 97074 W\"urzburg, Germany}}

\author{B. J. Ramshaw}
\affiliation{\mbox{Laboratory of Atomic and Solid State Physics, Cornell University, Ithaca, New York 14853, USA}}
\affiliation{Canadian Institute for Advanced Research, Toronto, Ontario, Canada}

\author{K. A. Modic}
\affiliation{\mbox{Institute of Science and Technology Austria, 3400 Klosterneuburg, Austria}}

\author{Fakher F. Assaad}
\affiliation{\mbox{Institut f\"ur Theoretische Physik und Astrophysik, Universit\"at W\"urzburg, 97074 W\"urzburg, Germany}}
\affiliation{\mbox{W\"urzburg-Dresden Cluster of Excellence ct.qmat, Germany}}

\begin{abstract}
 We compute the rotational anisotropy of the free energy of $\alpha$-RuCl$_3$ in an external magnetic field. This quantity, known as the magnetotropic susceptibility, $k$, relates to the second derivative of the free energy with respect to the angle of rotation.  We  have used approximation-free, auxiliary-field quantum Monte Carlo simulations for a realistic model of $\alpha$-RuCl$_3$  and  optimized  the  path integral to   alleviate  the  negative  sign problem. This  allows us  to  reach  temperatures  down to  $30~\rm{K}$---an energy scale below the dominant Kitaev  coupling.  
We demonstrate that the magnetotropic spin susceptibility in this model of $\alpha$-RuCl$_3$ displays scaling behavior $k = Tf(B/T)$ at high temperatures.
Once the uniform susceptibility departs from the Curie law (i.e., at the energy scale of the exchange interactions), it appears to transition to an emergent scalinglike behavior, characterized by a different function $f$ at lower temperatures, stemming from the locality of torque fluctuations.
We observe a remarkable  numerical  match  between experiment and  simulations and  we also find qualitative agreement with the pure Kitaev model. 
In comparison, for the XXZ Heisenberg Hamiltonian, the scaling $k = Tf(B/T)$ breaks down at a temperature scale where the uniform spin susceptibility deviates from the Curie law and never reemerges at low temperatures.
\end{abstract}

\maketitle

{\it Introduction.}
Quantum spin liquids are believed to harbor exotic fractionalized excitations that defy the conventional categories of fermions and bosons.
The Kitaev model, originally proposed by Kitaev in 2006~\cite{Kitaev06},  has served as a paradigm in this context, offering an exact solution for a quantum spin liquid state on the honeycomb lattice.

$\alpha$-RuCl$_3$ has emerged as a leading candidate for realizing the Kitaev spin liquid ~\cite{Jackeli-KSL-01,Chaloupka-2013, Plumb2014}.
Numerous experiments have probed its thermodynamic and dynamical properties and have reached the conclusion that 
there is a large Kitaev exchange interaction~\cite{Majumder2015,Sears2015,Kubota2015, Do:2017aa,Banerjee1055,Leahy17, Richard18, Kasahara2018,Lampen-Kelley2018,Banerjee:2018aa,Tanaka:2022aa}. One intriguing observation is the emergence of scale invariance at low temperatures and in high magnetic fields \cite{Modic20}. The aim of this Letter is to bridge this experimental observation in $\alpha$-RuCl$_3$ with approximation-free  finite temperature numerical  simulations. Using minimal models to describe $\alpha$-RuCl$_3$ we
 will see  that there is a  domain of  temperatures  and 
magnetic fields  where  numerics and  experiments  agree   quantitatively.    
With  the numerical  approach  we  can probe  higher  magnetic fields than  accessible in the laboratory, thereby suggesting the presence of two separate scaling regimes, each associated with a different scaling function.   
While the high temperature regime  is  generic  to all  spin models  and  emerges  at scales  well  above  the magnetic  exchange scale,  the lower scaling regime  is  characteristic of  materials  proximate  to the Kitaev   model.

\begin{figure}[t]                              
\begin{center}    
\centerline{\includegraphics[width=0.48\textwidth]{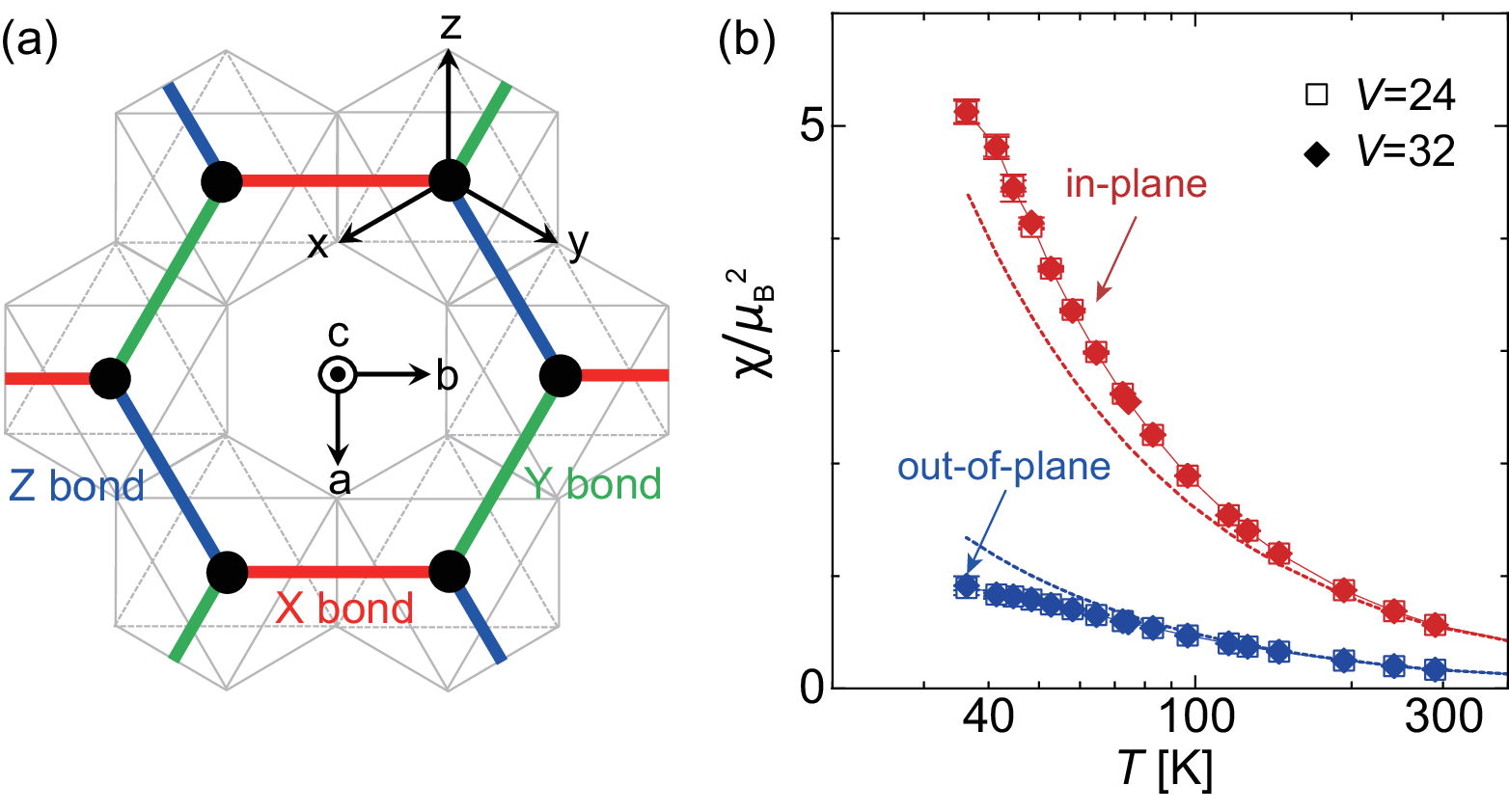}}
\caption[]{
 (a) Schematic of our model for $\alpha$-RuCl$_3$. 
Spin-1/2 degrees of freedom, $\hat{\boldsymbol{S}}_{\vec{i}}$, are situated on the honeycomb lattice and are subject to first-neighbor Kitaev $K_1$, and off-diagonal symmetric $\Gamma_1$, exchange interactions, as well as first-neighbor ($J_1$) and third-neighbor ($J_3$) Heisenberg couplings. 
The RuCl$_6$ octahedra within the honeycomb lattice, along with the definition of the cubic spin-space axes $[x,y,z]$ as indicated in the octahedron's top view, are illustrated.
(b) $T$ dependence of the in-plane and out-of-plane uniform spin susceptibilities $\chi$ from QMC simulations.
The dashed line is a fit to Curie's law at high temperatures.}
\label{fig:modelsus}                                     
\end{center}
\end{figure}

{\it Minimal models for Kitaev materials.}
We consider first-neighbor, Kitaev $K_1$, and off-diagonal symmetric, $\Gamma_1$, couplings, as well as first-neighbor (third-neighbor) Heisenberg couplings, $J_1$ ($J_3$), on the honeycomb lattice:
\begin{eqnarray}
\label{Eq:MMKM}
\hat{H}_s&=&\sum_{\ve{i}  \in A ,\gamma}\left[K_1\hat{S}_{\ve{i}}^{\gamma} \hat{S}_{\ve{i}+\ve{\delta}_\gamma}^{\gamma}+\Gamma_1 \left(\hat{S}_{\ve{i}}^{\alpha} \hat{S}_{\ve{i}+\ve{\delta}_\gamma}^{\beta}+\hat{S}_{\ve{i}}^{\beta} \hat{S}_{\ve{i}+\ve{\delta}_\gamma}^{\alpha}\right)\right]  \nonumber \\
& &+ \sum_{\ve{i}  \in A ,\ve{\delta}_\gamma}  J_{1}  \hat{\ve{S}}_{\ve{i}}  \cdot \hat{\ve{S}}_{\ve{i}+\ve{\delta}_\gamma} + \sum_{\ve{i} \in A ,\ve{\delta}'_\gamma}  J_{3}  \hat{\ve{S}}_{\ve{i}}  \cdot \hat{\ve{S}}_{\ve{i}+\ve{\delta}'_\gamma}.
\end{eqnarray}
Here $\ve{i}$ runs over the $A$ sublattice and $\ve{i}+\ve{\delta}_\gamma$ ($\ve{i}+\ve{\delta}'_\gamma$) over the first (third) neighbors.
For the first term $(\gamma,\alpha,\beta)=(1,2,3)$ for the $X$ bonds, $(\gamma,\alpha,\beta)=(2,3,1)$ for the $Y$ bonds, and $(\gamma,\alpha,\beta)=(3,1,2)$ for the $Z$ bonds on each lattice site [see Fig.~\ref{fig:modelsus}(a)]. 

To study the magnetotropic susceptibility of $\alpha$-RuCl$_3$ under high magnetic fields reported in Ref.~\cite{Modic20}, we add a Zeeman term to produce the total Hamiltonian
\begin{eqnarray}
\label{Eq:HKitaevplusB}
\hat{H}=\hat{H}_s-\mu_B\sum_{\ve{i}} \ve{B} \cdot  \hat{g} \cdot \hat{\ve{S}}_{\ve{i}},
\end{eqnarray}
where the direction of the magnetic field in the cubic spin basis corresponds to $\ve{B}||[xyz]$ [see Fig.~\ref{fig:modelsus}(a)].
In Kitaev materials such as $\alpha$-RuCl$_3$, the $[111]$ axis aligns with the $\boldsymbol{c}$ axis, perpendicular to the honeycomb lattice, whereas the $[11\bar{2}]$ and $[\bar{1}10]$ axes correspond to the in-plane $\boldsymbol{a}$ and $\boldsymbol{b}$ axes, respectively [see Fig.~\ref{fig:modelsus}(a)].
We adopt the parametrization $\ve{B}=B\left[\sin(\varphi)\sin(\theta)\ve{e}_{a}+\cos(\varphi)\sin(\theta)\ve{e}_{b}+\cos(\theta)\ve{e}_{c}\right] $, where the unit vectors $\ve{e}_{a}$, $\ve{e}_{b}$, and $\ve{e}_{c}$ point along the $[11\bar{2}]$, $[\bar{1}10]$, and $[111]$ directions, respectively.
$\hat{g}=g^{\alpha,\alpha'}$ represents the anisotropic $g$ factor, which contains only diagonal entries in the aforementioned directions, specifically  $(g_a, g_b,g_{c})=(2.3,2.3,1.3)$~\cite{Chaloupka16,Yadav:2016aa}.  

There are  many methods, such as finite temperature  Lanczos \cite{Winter18} or thermal  pure  quantum states \cite{Suzuki18,Laurell20},  to  simulate finite   temperature  properties  of  the  aforementioned  Hamiltonian. 
 Here we opt for  quantum Monte Carlo (QMC) simulations that excel at  thermodynamic quantities.
Frustrated  spin systems, like Kitaev materials, generically suffer from the infamous negative sign problem that leads to an exponential increase in the required computational power as a function of the volume $V$ of the system and inverse temperature $\beta$~\cite{Troyer05}.
Since the severity of this problem depends on the specific formulations, optimization strategies to alleviate it can be put forward.
Indeed, in our recent publication~\cite{SatoT20_1}, we have developed a fermion QMC approach using the auxiliary-field QMC (AFQMC) 
 algorithm for fermions~\cite{Blankenbecler81,White89,ALF_v1,ALF_v2} to tackle frustrated spin models.
The generalized Kitaev model, describing materials, such as layered iridates and $\alpha$-RuCl$_3$, benefits from this approach. It mitigates the severity of the negative sign problem and enables QMC simulations at temperatures well below the magnetic exchange scale. 
This opens up a window of temperatures relevant to experiments.
We demonstrate that this method reproduces the experimental magnetotropic susceptibility in $\alpha$-RuCl$_3$~\cite{Modic20}.
 Hamiltonian ($\ref{Eq:MMKM}$) was simulated using  the AFQMC method of Ref.~\cite{SatoT20_1} and we refer the reader to  this paper  for details of  the approach.
 We use  a  Trotter discretization in  the  range  $\Delta\tau  \in [0.01, 0.05]$ depending upon the temperature.  
For this range of  $\Delta\tau $, the systematic error is contained within our error bars.  
We simulated lattices with $L \times L$ unit cells (each containing two spins, i.e., $V = 2L^2$ lattice sites on the honeycomb lattice) and periodic boundary conditions.

{\it Uniform spin susceptibilities.}
A key question is whether our QMC approach allows one to reach temperature scales that are relevant to experiments for Kitaev materials. 
To determine the lowest accessible temperature, we numerically evaluate  the spin susceptibility tensor $\hat{\chi}=\chi^{\alpha,\alpha'}$,
\begin{eqnarray}
\label{eq:spinsus}
\chi^{\alpha,\alpha'}(\vec{q})=  \int_{0}^{\beta} \text{d} \tau   \Big[
\langle  \hat{\vec{O}}^{\alpha}_{\vec{q}}(\tau)  \hat{\vec{O}}^{\alpha'}_{-\vec{q}}(0) \rangle-\langle  \hat{\vec{O}}^{\alpha}_{\vec{q}}  \rangle \langle  \hat{\vec{O}}^{\alpha'}_{-\vec{q}} \rangle \Big]
\end{eqnarray}
where $ \hat{\vec{O}}^{\alpha}_{\vec{q}}  = \frac{\mu_B g^{\alpha,\alpha}}{\sqrt{V}} \sum_{\vec{r}} e^{i  \vec{q} \cdot \vec{r}} \left( \hat{S}_{\vec{r},A}^{\alpha}+\hat{S}_{\vec{r},B}^{\alpha}e^{i\vec{q}\vec{R}} \right)$.
Here $\ve{r}$ runs over the $A$ sublattice (or unit cell)  and $\vec{R}=2/3( \ve{A}_2-\ve{A}_{1}/2)$ with the primitive lattice vectors {$\ve{A}_1=(1,0)$ and $\ve{A}_2=(\frac{1}{2},\frac{\sqrt{3}}{2})$.}
Projecting $\chi^{\alpha,\alpha'}(\ve{q=\Gamma})$ onto the in-plane (out-of-plane) direction yields the in-plane (out-of-plane) uniform spin susceptibility 
$\chi_{\parallel}=\ve{e}_{ab}^{\text{T}}\hat{\chi}\ve{e}_{ab}$ ($\chi_{\perp}=\ve{e}_{c}^{\text{T}}\hat{\chi}\ve{e}_{c}$). 

Figure~\ref{fig:modelsus}(b) plots the results down to the lowest accessible  temperature. 
For model parameters proposed to describe $\alpha$-RuCl$_3$~\cite{Winter:2017aa},
 $(J_1, J_3, K_1, \Gamma_1)=(-5.8,5.8,-58,29)~{\rm K}$, we can reach temperatures down to $T\sim 30~\rm{K}$ on the relatively large lattice size $V= 32$, which is larger than the  lattice sizes  in exact diagonalization calculations at finite temperature (i.e., $V=24$ sites~\cite{Winter18}).
On the experimental front, $\alpha$-RuCl$_3$ exhibits zigzag spin order at low temperatures, but proximity to the Kitaev spin liquid suggests that high energy features of this material are described by Majorana fermions~\cite{Do:2017aa,Banerjee1055}.
These fermions will hence only show up in finite-temperature properties in an intermediate temperature range bounded by the ordering temperature from below and the coherence scale of the Majorana fermions from above.   Experimentally,  this  temperature  range   corresponds  to $ T \in [10,100 ]~\rm{K}$~\cite{Do:2017aa,Banerjee1055},   and  it  is  remarkable  to  observe  that  the  QMC simulations  can    access  this regime  before  the  negative  sign  problem 
becomes too  severe. 
Our numerical results, as a function of decreasing temperature, not only confirm the deviation from a Curie law at high temperatures but also demonstrate a clear trend in $\chi_{\parallel}>\chi_{\perp}$, which exhibits similar behavior as experiments in $\alpha$-RuCl$_3$ reported in Refs.~\cite{Majumder2015,Sears2015,Kubota2015,Lampen-Kelley2018}.

\begin{figure}[t]
\centering
\centerline{\includegraphics[width=\columnwidth,trim={0cm 0 0.5cm 0},clip]{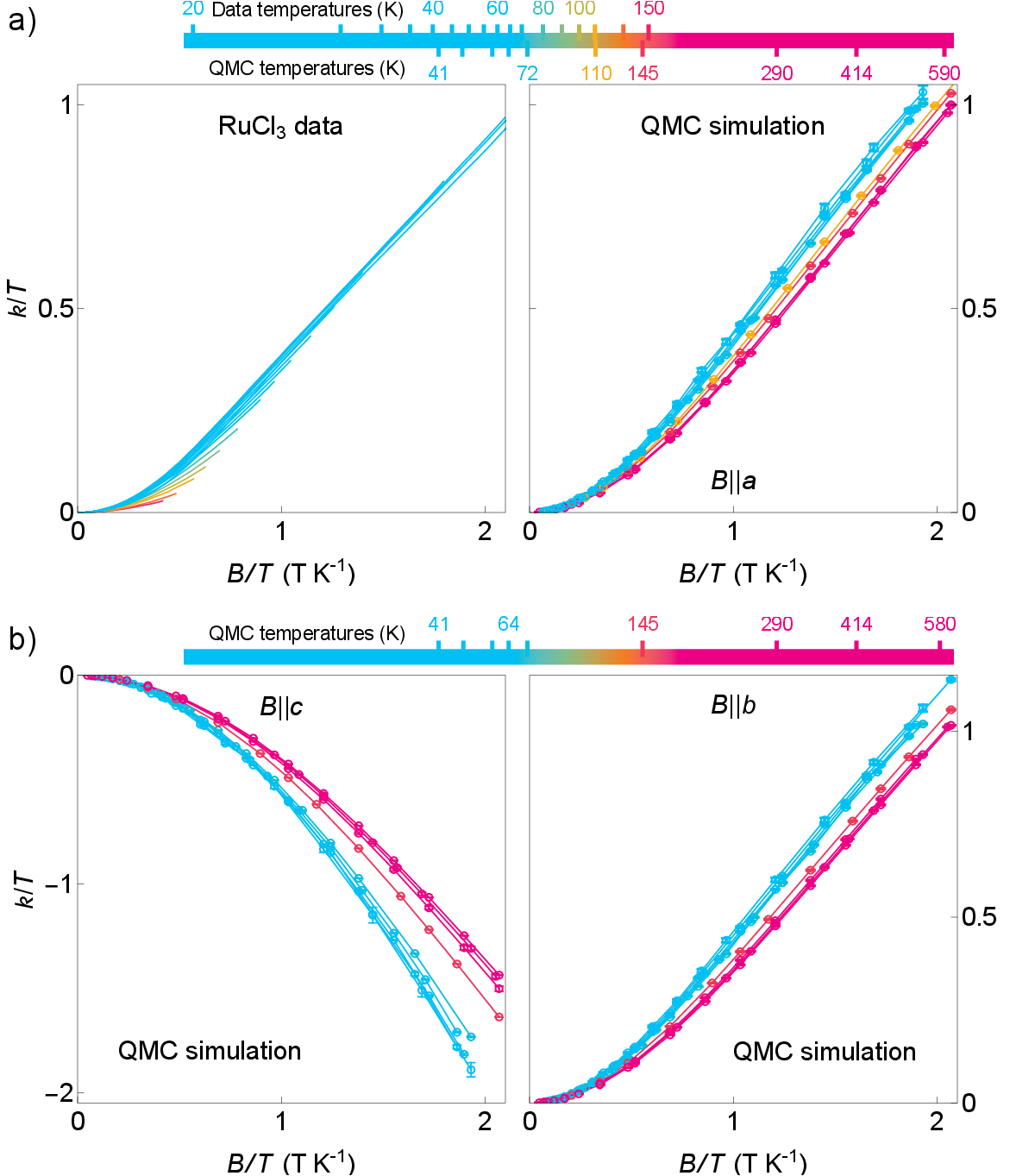}}
\caption{\label{fig:MC-1} 
(a) The magnetotropic susceptibility per Ru atom of $\alpha$-RuCl$_3$ normalized by temperature $k/T$ versus magnetic field normalized by temperature $B/T$. Both $k$ and $T$ are given in energy units for the vertical axis, making $k/T$ a dimensionless quantity. $T$ is in kelvin for the lower axis. Left panel a shows the experimental data from~\cite{Modic20} and right panel shows our QMC calculations using the model parameters described in the text. The applied magnetic field is oriented close to the $\ve{a}$-$\ve{b}$ plane. The collapse of blue curves in both the experimental data and the calculated magnetotropic susceptibility appear to show scaling behavior in the temperature range $ T \in [20,70]~\rm{K}$. Above this temperature, the experimental data deviates toward the paramagnetic scaling (red curves) observed at high temperatures in the QMC calculations. Limited field range at these temperatures prevents experimental access to the high-temperature scaling regime. 
(b) Same as (a) for varying magnetic field directions in the QMC calculations. }
\end{figure}

{\it Magnetotropic susceptibility.}
We now focus our attention on recent measurements for $\alpha$-RuCl$_3$ concerning the  
magnetotropic susceptibility over a wide range of temperatures and magnetic fields~\cite{Modic20}.   This quantity is defined as the second derivative of the free energy with respect to the rotation angle of the magnetic field and is the thermodynamic coefficient associated with the magnetic anisotropy.  
In our QMC simulations, and as detailed in the Supplemental Material~\cite{sup}, the magnetotropic susceptibility in the rotation plane perpendicular to the unit vector $\ve{e}$ is computed using  \cite{Mandal21,Shekhter23}
\begin{eqnarray}
  \label{MC}
k&=&  \frac{1}{V}  \Big[ \mu_{B} \ve{e} \times \left( \ve{e} \times \ve{B} \right) \cdot \hat{g} \cdot \langle   \hat{\ve{S}}_{tot} \rangle\nonumber\\ 
&-&\mu_{B}^2
 \int_0^{\beta} d\tau  \, \ 
  \Big[
  \langle
    \left( \ve{e} \times \ve{B} \right) \cdot \hat{g} \cdot  \hat{\ve{S}}_{tot}(\tau) 
     \left( \ve{e} \times \ve{B} \right) \cdot \hat{g} \cdot  \hat{\ve{S}}_{tot}(0)
  \rangle \nonumber\\ 
  &&~~~~ -
  \langle
   \left( \ve{e} \times \ve{B} \right) \cdot \hat{g} \cdot  \hat{\ve{S}}_{tot} 
     \rangle^2   \Big) \Big]
\end{eqnarray}
with $\hat{\boldsymbol{S}}_{tot}  = \sum_{\ve{i}}\hat{\ve{S}}_{\ve{i}}$ and an imaginary time $\tau$.
Figure~\ref{fig:MC-1} presents the results for the magnetotropic susceptibility $k$, with respect to temperature and magnetic field in the $\ve{a}$-$\ve{b}$ plane, and compares them to experiments for $\alpha$-RuCl$_3$~\cite{Modic20}.
The magnetotropic susceptibility measurements appear to show temperature and magnetic field scaling behavior of the form $k=Tf(B/T)$ over the temperature range bounded by the spin ordering temperature and the coherence scale of Majorana fermions, $T \approx [10,100 ]~\rm{K}$ [see the left panel of Fig.~\ref{fig:MC-1}(a)].
Our QMC data shown in the right panel of Fig.~\ref{fig:MC-1}(a) quantitatively reproduces this behavior below the magnetic exchange scale.
Furthermore, our numerical results suggest that as the temperature increases, there is a departure from the low-temperature behavior, and we then observe a more well-defined scaling behavior at high temperatures. 
The experimental data shows the departure from the low-temperature scalinglike behavior, but higher magnetic fields are required to reach the high-temperature, paramagnetic scaling.
Moreover, for the other magnetic field directions in the QMC data, for instance, along the $\ve{c}$ direction [the left panel of Fig.~\ref{fig:MC-1}(b)] and the $\ve{b}$ direction [the right panel of Fig.~\ref{fig:MC-1}(b)], the observed behavior remains consistent at both high and low temperatures.

Scaling behavior of the form $k=Tf(B/T)$ is  satisfied  for independent  local  moments  for any anisotropic $g$-factor (see Supplemental Material~\cite{sup}). This is expected for any spin system when the temperature exceeds the magnetic exchange energy and the scaling is expected to break down below this energy scale.
However, the case of $\alpha$-RuCl$_3$ reveals a more complex scenario. 
Our QMC data show two behaviors: scaling that is characteristic of a free spin at high temperatures, and an emergent behavior at low temperatures. 

\begin{figure}[b]
\centering
\centerline{\includegraphics[width=0.48\textwidth]{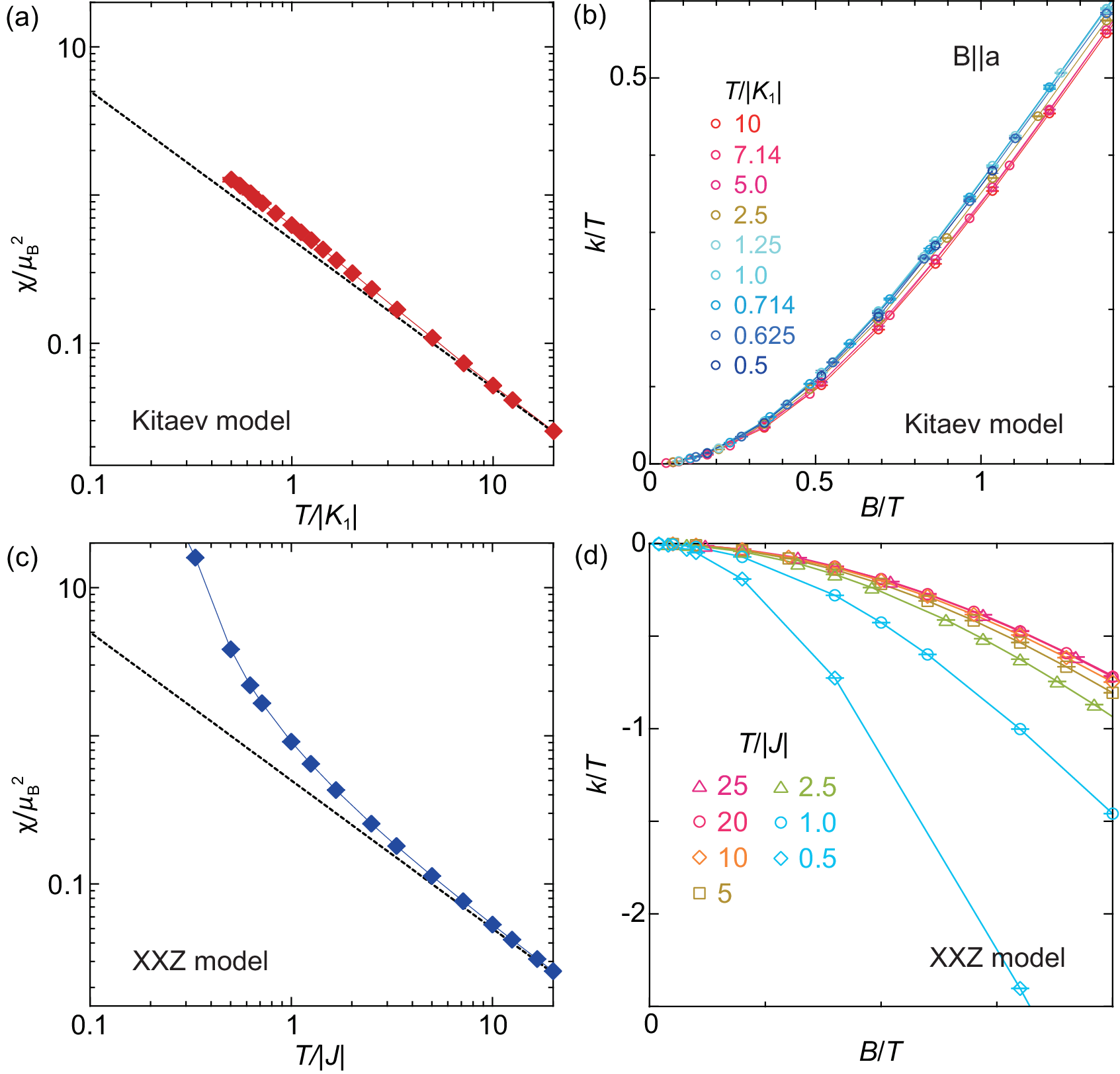}}
\caption{\label{fig:MC-3} 
$T$ dependence of the uniform spin susceptibilities $\chi$ (left panels) and the magnetotropic susceptibility normalized by temperature $k/T$ versus magnetic field  normalized by temperature $B/T$ (right panels).
(a), (b) The Kitaev model with $\Gamma_1 =  J_1  =  J_3 = 0 $ in our model for $\alpha$-RuCl$_3$, and (c), (d) the nonfrustrated spin model on the honeycomb lattice with $(J,J_z)=(-1,0.5)$ and $V= 72$ considered the out-of-plane magnetic-field direction (see the main text).
The dashed line represents a fit to Curie's law at high temperatures.
}
\label{fig:unisus} 
\end{figure}

To  underline the  uniqueness of the low-temperature behavior of the magnetotropic susceptibility observed in $\alpha$-RuCl$_3$, we now compare our findings  with other models.
Let us  start  with the pure Kitaev Hamiltonian,  as  obtained  by  setting $\Gamma_1 =  J_1  =  J_3 = 0 $ in Eq.~(\ref{Eq:MMKM}) and  using  the 
same magnetic field  orientation as in Fig.~\ref{fig:MC-1}(a) [see Figs.~\ref{fig:MC-3}(a) and \ref{fig:MC-3}(b)]. 
It is  remarkable to see that we  observe  the very same  behavior: a well-defined scaling at high temperatures and an emergent low-temperature one,  albeit with different numerical  values.  
The  temperature  at  which  we  observe the  crossover between the high and low  temperature behavior matches the   one   at  which the uniform spin susceptibility    departs  from the  Curie law. 
This result confirms  the  notion that $\alpha$-RuCl$_3$ is  proximate  to the Kitaev model, and  that the  low temperature behavior is a  distinct  property  of  the  Kitaev  model. 

We now concentrate on a nonfrustrated spin model.
To this end we  consider  for  the  XXZ  model on the honeycomb lattice, 
$\hat{H}_s=\sum_{\langle \ve{i}, \ve{j} \rangle}  J \left[  \hat{S}^x_{\ve{i}}  \cdot \hat{S}^x_{\ve{j}} +  \hat{S}^y_{\ve{i}}  \cdot \hat{S}^y_{\ve{j}} \right]
+ \left[  J + J_{z} \right] \hat{S}_{\ve{i}}^{z} \hat{S}_{\ve{j}}^{z} $, 
 and present QMC results for $J_z/J = -0.5$ in Fig.~\ref{fig:MC-3}. 
As the  temperature decreases, the uniform spin susceptibility $\chi$ deviates from the high-temperature Curie law [see Fig.~\ref{fig:MC-3}(c)].
In the ferromagnetic case, $J=-1$, $\chi$ grows and ultimately diverges at low temperatures.
As is apparent from the data in Fig.~\ref{fig:MC-3}(d), our numerical results for  the  magnetotropic susceptibility confirm  the  high-temperature scaling behavior:  all curves for $T/|J| > 10$  collapse when scaled in this way. 
The  breakdown  of  the data  collapse  $k=Tf(B/T)$ agrees with the temperature scale where the deviation from the Curie law behavior is observed in the susceptibility, and the low-temperature data shows no further collapse into a new scaling form.
Note that we have checked that the antiferromagnetic case, $J=1$, also exhibits no such recollapse at low temperatures (see Supplemental Material~\cite{sup}).

\begin{figure}[t]
\centering
\centerline{\includegraphics[width=0.48\textwidth]{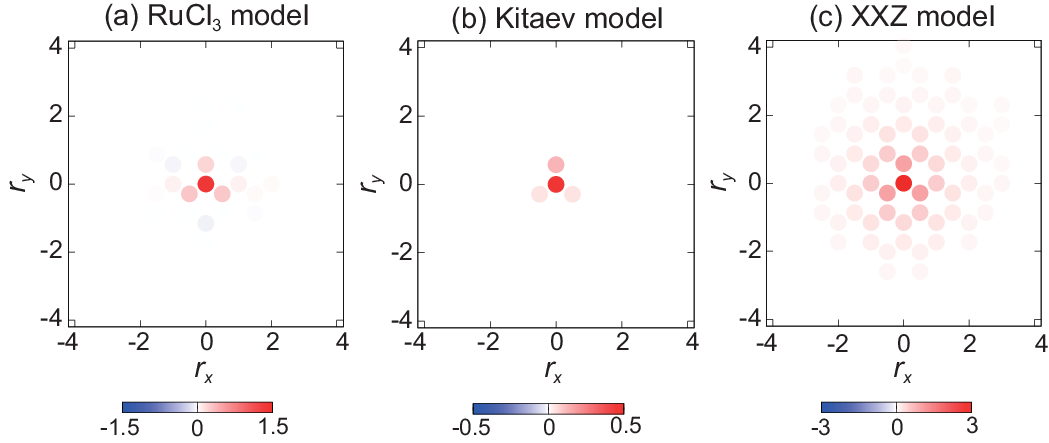}}
\caption{\label{fig:torque} 
Real-space torque correlations $ \langle  t_{\ve{r}} t_{\ve{0}}  \rangle =  \sum_{\alpha,\gamma}  b_{\alpha} b_{\gamma}   \left< \hat{S}_{\ve{r}}^{\alpha} \hat{S}_{\ve{0}}^{\beta}   \right>    $   with $\ve{b} = (  \ve{e} \times \ve{B}   ) \cdot \hat{g}   $   for 
the RuCl$_3$ model (a) [$T=41.4~\rm{K}$ and $B=5~\rm{T}$]
and
the Kitaev model (b) [$\Gamma_1 =  J_1  =  J_3 = 0 $ in our model for $\alpha$-RuCl$_3$, $T/|K_1|=0.5$ and $\mu_B B/|K_1|=0.0579$]
and
the XXZ model with $(J,J_z)=(-1,0.5)$ (c) [$T/|J|=0.5$ and $\mu_B B/|J|=0.1$].
Here, the magnetic field is considered in the out-of-plane direction. 
Note that in the high temperature   scaling regime  where the temperature  is  
larger  than the  magnetic  exchange coupling,  the  torque correlations are purely  local. 
}
\label{fig:torque} 
\end{figure}

{\it Summary and  interpretation.}
We have investigated the behavior of the magnetotropic susceptibility, $k$, in a realistic  magnetic model of $\alpha$-RuCl$_3$ in an external magnetic field.
Employing approximation-free, auxiliary-field quantum Monte Carlo simulations, we have explored the rotational anisotropy of the free energy, highlighting the intricate relationship between temperature $T$, magnetic field $B$, and the unique scaling behavior of $k$.
Our results show, along with a well-defined scaling behavior at high temperatures, $k=Tf(B/T)$, an emergent scalinglike behavior at low temperatures, characterized by a different function, $f$.
The low-temperature behavior is in quantitative agreement with the experimental data~\cite{Modic20} (Fig.~\ref{fig:MC-1}). 
The high-temperature  scaling, which is inaccessible to experiments, is generic to all spin  systems: when the thermal  energy  exceeds  the  magnetic  exchange, spin systems behave as independent  local  moments.
The low-temperature  behavior, on the other hand, is observed in our model of  $\alpha$-RuCl$_3$ as well as in the Kitaev Hamiltonian and is not present for the unfrustrated XXZ model.

The    magnetotropic susceptibility   relates to  the  fluctuations  of  the  torque  $ \frac{\partial F} {\partial \theta} =   \mu_B \sum_{\ve{i}} \hat{t}_{\ve{i}} $  with $ \hat{t}_{\vec{i}} = (  \ve{e} \times \ve{B}   ) \cdot \hat{g}  \cdot \hat{\ve{S}}_{\ve{i} }$. 
A naive explanation for the emergence of scaling at low temperature would be proximity to a critical point  at  which   long   wave length   fluctuations of  the  torque, $\hat{t}_{\ve{i}}$,   become  critical  and   dominate    the  value  of  $k$.      In  Fig.~\ref{fig:torque}    we  plot  the  real  space   fluctuations  of   $\hat{t}_{\ve{i}}$   both  for  the  model  of  $\alpha$-RuCl$_3$  at  our  lowest   temperature   and  for   the XXZ  model.      While   torque  fluctuations  build  up    for  the  XXZ  model,  they  remain  very  short  ranged   for   $\alpha$-RuCl$_3$ and for  the  Kitaev model
\footnote{Note that at  temperatures  beyond  the magnetic exchange,   torque correlations vanish between any two distinct  lattice sites.}.  
As a consequence, the low-temperature data collapse cannot be understood within a renormalization group approach in which scaling behavior stems from a low-energy effective theory in the vicinity of a critical point.  

Instead  of  being  the  signature  of   long-ranged  critical   torque  fluctuations,    the  observed  behavior at low temperatures  can   be  interpreted  in terms  of  the  
\textit{absence}  of  such  fluctuations  beyond  the length scale  set  by   the  lattice  constant    and  down  to   temperature  scales  $T=20~{\rm K}$  corresponding   to  a third  of  the   dominant  Kitaev  coupling.  As  a  consequence,  the  adequate  model   to   understand this behavior   is   that of  a renormalized local  moment,   satisfying  $k = Tf(B/T)$.    Corrections to  this  model   can  not  be  argued  away  with e.g. a    renormalization group   argument,   and  we  understand that    the  scaling    has  to be  approximate. 

This  point  of  view  places    strong  constraints  on the   low  lying  excitations  of  Kitaev materials: 
the  magnetotropic susceptibility  captures only high  energy local correlations,  while the  low  energy  excitations 
are magnetically isotropic  and  hence do not  contribute  to the magnetotropic  susceptibility. The  qualitative agreement between experimental 
and  numerical  data for  the   Kitaev  Hamiltonian supports this point of view. 
In the  Supplemental  Material~\cite{sup} we  show  that   
one  can formulate,  for  the Kitaev  model,   a  mean-field   theory   in  which  the  low lying  excitations  have  no  magnetic  anisotropy    and  hence have vanishing    torque    fluctuations.    

Our results lead to a flurry of  further  applications.   We can refine the parameters  of  the model by  fitting  numerics  to the vast amount   angle  and 
temperature  dependent data.  We  have merely looked into the shift of  the resonant frequency,  corresponding to the  real part of  the  magnetotropic susceptibility.   The line-width picks up  the energy  imaginary part of  the  response function,  that should  show  unique  behavior at, e.g., quantum phase  transitions.   Finally,  a  host of  other magnetic and metallic  materials can and should be investigated with this approach.

\bigskip
\begin{acknowledgments}
We  gratefully acknowledge the Gauss Centre for Supercomputing e.V. for funding this project by providing computing time on the GCS Supercomputer SUPERMUC-NG at the Leibniz Supercomputing Centre  (project No. pn73xu)
as  well  as  the scientific support and HPC resources provided by the Erlangen National High Performance Computing Center (NHR@FAU) of the Friedrich-Alexander-Universit\"at Erlangen-N\"urnberg (FAU) under the NHR Project b133ae. NHR funding is provided by federal and Bavarian state authorities. NHR@FAU hardware is partially funded by the German Research Foundation (DFG) -- 440719683.   
T.S. thanks funding from the Deutsche Forschungsgemeinschaft under the Grant No. SA 3986/1-1  as well as the W\"urzburg-Dresden Cluster of Excellence on Complexity and Topology in Quantum Matter ct.qmat (EXC 2147, Project ID 390858490).   F.F.A.    acknowledges financial support from the German Research Foundation (DFG) under the grant AS 120/16-1 (Project No. 493886309) that is part of the collaborative research project SFB Q-M\&S funded by the Austrian Science Fund (FWF) F 86.
K.A.M. thanks financial support from the  Austrian Science Fund,  SFB F 86, Q-M\&S.
\end{acknowledgments}

%

\clearpage
\section*{Supplementary information}

\subsection*{magnetotropic susceptibility}
The magnetotropic susceptibility quantifies how the free energy of a system changes as the direction of the magnetic field varies, especially under small rotations.
This quantity, denoted as $k$, is defined as the second derivative of the free energy with respect to the rotation angle $ \lambda$ of the magnetic field
\begin{eqnarray}
\label{Eq:SM-MC-0}
k=\frac{\partial^2 F}{\partial \lambda^2}\bigg|_{\lambda=0}.
\end{eqnarray}
Now consider an arbitrary vector $\ve{B}$ representing the magnetic field. 
Assume that the direction of this vector fluctuates within a plane. 
This plane is defined by its normal vector $\ve{e}$. 
The fluctuations or oscillations occur around the direction of $\ve{B}$, within the defined plane.
To describe these rotations, we employ the generators of SO(3) symmetry, specifically $e^{i \hat{\ve{K}}\cdot \ve{e} \lambda}$.
Here, $\hat{\ve{K}}$ consists of purely imaginary entities
\begin{eqnarray}
\label{Eq:SM-MC-1}
\hat{K}_1 &=&
    \left[
    \begin{array}{ccc}
    0 & 0 & 0 \\
    0 & 0 & -i\\
    0 & i & 0 
    \end{array}
    \right]~~,~~
    \hat{K}_2 =
    \left[
    \begin{array}{ccc}
    0 & 0 & i \\
    0 & 0 & 0 \\
    -i & 0 & 0
    \end{array}
    \right] \nonumber\\ 
    \hat{K}_3 &=&
    \left[
    \begin{array}{ccc}
    0 & -i & 0 \\
    i & 0 & 0\\
    0 & 0 & 0 
    \end{array}
    \right]
\end{eqnarray}
which satisfy  the commutation relations
$[\hat{K}_{\alpha}, \hat{K}_{\beta}]=i \epsilon_{\alpha,\beta,\gamma}\hat{K}_{\gamma}$.
Within this, when the magnetic field is rotated by small angle $\lambda$, the total Hamiltonian from the main text is described as
\begin{eqnarray}
\label{Eq:SM-MC-2}
\hat{H}(\lambda)=\hat{H}_s-\mu_B e^{i \hat{\ve{K}}\cdot \ve{e} \lambda}\ve{B} \cdot  \hat{g} \cdot \hat{\ve{S}}_{tot} 
\end{eqnarray}
with $\hat{\boldsymbol{S}}_{tot}  = \sum_{\ve{i},\alpha}\hat{S}_{\ve{i}}^{\alpha}$.
$\hat{g}$ represents the anisotropic $g$ factor.

The free energy, in relation to the Hamiltonian (\ref{Eq:SM-MC-2}), is given by
$F(\lambda)=-\frac{1}{\beta}\log   \text{Tr} \left[   e^{-\beta\hat{H}(\lambda) } \right] $ with the inverse temperature $\beta$.
We can now expand the Hamiltonian (\ref{Eq:SM-MC-2}) in terms of small $\lambda$, giving
\begin{eqnarray}
\label{Eq:SM-MC-3}
\hat{H}(\lambda)=\hat{H}_s+\hat{H}_1(\lambda)+O(\lambda^2).
\end{eqnarray}
Here, the term $\hat{H}_1$ captures the second-order effects of the rotation, and is given by
\begin{eqnarray}
\label{Eq:SM-MC-4}
\hat{H}_1=-\mu_B  \Big[ i\hat{\ve{K}}\cdot \ve{e}\lambda+\frac{(i\hat{\ve{K}}\cdot \ve{e})^2}{2}\lambda^2 \Big]\ve{B} \cdot  \hat{g} \cdot \hat{\ve{S}}_{tot}. \nonumber\\
\end{eqnarray}
This expansion allows to express the second derivative of the free energy with respect to the rotation angle $\lambda$ in terms of $\hat{H}_1$
\begin{eqnarray}
\label{Eq:SM-MC-5}
\frac{\partial^2 F}{\partial \lambda^2}\bigg|_{\lambda=0}
&=&\langle \frac{\partial^2 \hat{H}_1}{\partial\lambda^2}  \rangle \nonumber\\ 
&-&
 \int_0^{\beta} d\tau  \, \ 
  \Big[
  \langle \frac{\partial \hat{H}_1 (\tau) }{\partial \lambda} \frac{\partial \hat{H}_1 (0) }{\partial \lambda} \rangle 
  -\langle \frac{\partial \hat{H}_1}{\partial \lambda} \rangle\langle \frac{\partial \hat{H}_1}{\partial \lambda} \rangle 
    \Big].
  \nonumber\\ 
\end{eqnarray}
The integration runs over an imaginary time $\tau$.
Taking the first and second derivatives of $\hat{H}_1$ in Eq.~(\ref{Eq:SM-MC-4}), we obtain
\begin{eqnarray}
\label{Eq:SM-MC-6}
 \frac{\partial \hat{H}_1}{\partial \lambda}\bigg|_{\lambda=0}  &=& -\mu_{B} (i\hat{\ve{K}}\cdot \ve{e})\cdot\ve{B} \cdot \hat{g} \cdot \hat{\ve{S}}_{tot} \nonumber\\ 
 &=& \mu_{B} (\ve{e} \times \ve{B} )\cdot \hat{g} \cdot \hat{\ve{S}}_{tot}, \nonumber\\ 
\frac{\partial^2 \hat{H}_1}{\partial \lambda^2}\bigg|_{\lambda=0}  &=& \mu_{B} (i\hat{\ve{K}}\cdot \ve{e})^2 \cdot \ve{B} \cdot \hat{g} \cdot \hat{\ve{S}}_{tot} \nonumber\\ 
 &=& \mu_{B} \ve{e} \times (\ve{e} \times \ve{B} )\cdot \hat{g} \cdot \hat{\ve{S}}_{tot}.
\end{eqnarray}
Synthesizing Eqs.~(\ref{Eq:SM-MC-5}) and (\ref{Eq:SM-MC-6}), the expression for the magnetotropic susceptibility $k$ is
\begin{eqnarray}
k&=&  \frac{1}{V}  \Big[ \mu_{B} \ve{e} \times \left( \ve{e} \times \ve{B} \right) \cdot \hat{g} \cdot \langle   \hat{\ve{S}}_{tot} \rangle\nonumber\\ 
&-&\mu_{B}^2
 \int_0^{\beta} d\tau  \, \ 
  \Big[
  \langle
    \left( \ve{e} \times \ve{B} \right) \cdot \hat{g} \cdot  \hat{\ve{S}}_{tot}(\tau) 
     \left( \ve{e} \times \ve{B} \right) \cdot \hat{g} \cdot  \hat{\ve{S}}_{tot}(0)
  \rangle \nonumber\\ 
  &&~~~~ -
  \langle
   \left( \ve{e} \times \ve{B} \right) \cdot \hat{g} \cdot  \hat{\ve{S}}_{tot} 
     \rangle^2   \Big) \Big]
\end{eqnarray}     
with lattice sites $V$.
Within our QMC simulations, this form is employed to compute the magnetotropic susceptibility.

\subsection*{Properties of the magnetotropic susceptibility in free spin systems}
The main text references the temperature-magnetic field scaling behavior of the form $\beta k=f(\beta B)$ for free spins.
Now consider $\hat{H}_s=0$ in the Hamiltonian of Eq.~(\ref{Eq:SM-MC-2}).
In this case, the system is dominated solely by the external magnetic field and temperature. 
The corresponding free energy is then described by
\begin{eqnarray}
\label{Eq:SM-MCfree-0}
F(\lambda)&=&-\frac{1}{\beta}\log   \text{Tr} \left[    e^{\beta B \mu_B R(\ve{e}, \lambda)\ve{n} \cdot  \hat{g} \cdot \hat{\ve{S}}_{tot} } \right] \nonumber\\ 
&=&-\frac{1}{\beta}g(\beta B, \lambda).
\end{eqnarray}
Here, $R(\ve{e}, \lambda)=e^{i \hat{\ve{K}}\cdot \ve{e} \lambda}$ represents the rotation matrix in the presence of the magnetic field, and $\ve{B}=B\ve{n}$ with $\ve{n}$ being the unit vector pointing in the direction of the magnetic field. 
This simplified expression for the free energy provides an explicit link between the rotation angle and the behavior of the system under a magnetic field.
Taking this forward, the magnetotropic susceptibility $k$ in this context is given by
\begin{eqnarray}
\label{Eq:SM-MCfree-1}
k&=&\frac{\partial^2 F(\lambda)}{\partial \lambda^2}\bigg|_{\lambda=0} \nonumber\\ 
&=&-\frac{1}{\beta}\frac{\partial^2 g(\beta B, \lambda)}{\partial \lambda^2}\bigg|_{\lambda=0}.
\end{eqnarray}
Upon further simplification, we arrive at the key relationship:
\begin{eqnarray}
\label{Eq:SM-MCfree-2}
\beta k=f(\beta B).
\end{eqnarray}
This equation accentuates the intricate relationship between the temperature and magnetic field strength.  

\begin{figure}[t]
\centering
\centerline{\includegraphics[width=0.48\textwidth]{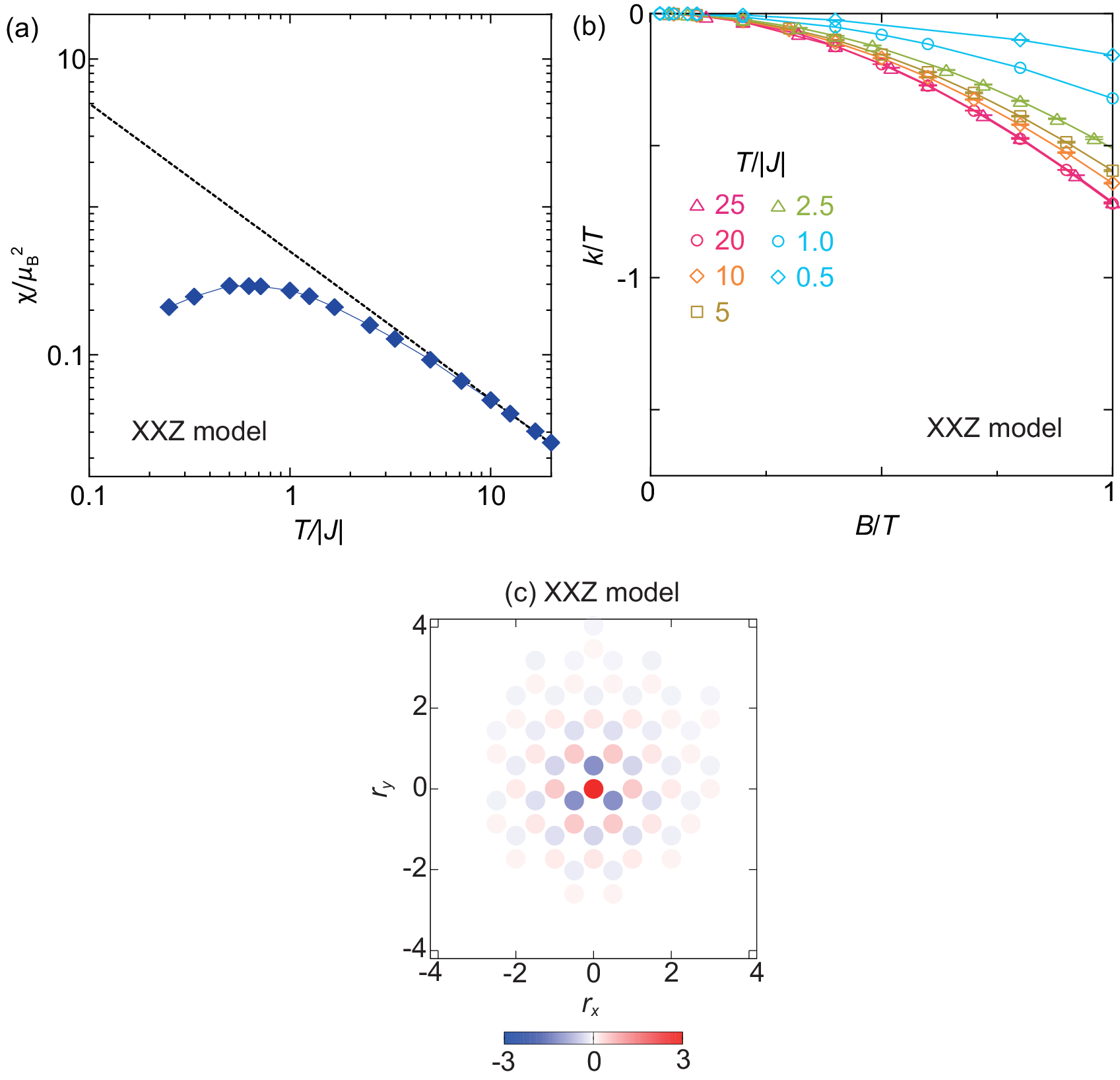}}
\caption{\label{fig:AFMXXZ-MC} 
QMC results for the XXZ model on the honeycomb lattice with $(J,J_z)=(1,-0.5)$.
(a) $T$ dependence of the uniform spin susceptibilities $\chi$. 
The dashed line represents a fit to Curie's law at high temperatures.
(b) The magnetotropic susceptibility normalized by temperature $k/T$ versus magnetic field  normalized by temperature $B/T$ and (c) real-space torque correlations  $ \langle  t_{\ve{r}} t_{\ve{0}}  \rangle$ [$T/|J|=0.5$ and $\mu_B B/|J|=0.1$].
The magnetic field is considered in the out-of-plane direction.
Here, $V=72$ lattice.
}
\label{fig:AFMXXZ-MC} 
\end{figure}

\subsection*{Magnetotropic susceptibility for the XXZ model}

In the main text, we consider the ferromagnetic case of the XXZ model on the honeycomb lattice. 
Now QMC results for the antiferromagnetic case, ($J,J_z)=(1,-0.5)$, are presented in Fig.~\ref{fig:AFMXXZ-MC}.
As the temperature decreases, the uniform spin susceptibility, $\chi$, deviates from the high-temperature Curie law (see Fig.~\ref{fig:AFMXXZ-MC}(a)).
Local antiferromagnetic correlations lead to a suppression of $\chi$ with respect to the high-temperature Curie law, and at low temperatures, $\chi$ will scale  to a constant value, reflecting the presence of Goldstone modes.
Figure~\ref{fig:AFMXXZ-MC}(b) shows the magnetotropic susceptibility $k$.
It shares the same features as for the ferromagnetic case shown in the main text.
Our numerical data confirm the data collapse $k=Tf(B/T)$ for $T/|J| > 10$ and its breakdown agrees with the temperature scale where the deviation from Curie law behavior is observed in $\chi$. 
Figure~\ref{fig:AFMXXZ-MC}(c) plots the  torque correlations.  
As is apparent, torque  fluctuations  build  up.

\subsection*{Abrikosov fermion representation and mean-field ansatz for the Kitaev model}
The Kitaev model consists of $S=1/2$ spins on a honeycomb lattice with Hamiltonian
\begin{eqnarray}
\label{Eq:SM-KMF-0}
\hat{H}=K_1\sum_{\ve{i}  \in A, \alpha} \hat{S}_{\ve{i}}^{\alpha} \hat{S}_{\ve{i}+\ve{\delta}_\alpha}^{\alpha}.
\end{eqnarray}
Here $\ve{i}$ runs over the A sublattice and $\ve{i}+\ve{\delta}_\alpha$ with $\alpha=(1,2,3)$ over the first neighbors.
Now we represent the spin-1/2 degree of freedom $\hat{S}_{\ve{i}}^{\alpha}$ in terms of Abrikosov fermions
\begin{eqnarray}
\label{Eq:SM-KMF-1}
	\hat{S}_{\ve{i}}^{\alpha} =\frac{1}{2} \sum_{s,s'}\hat{f}^{\dagger}_{\ve{i},s} \sigma^{\alpha}_{s,s' }   \hat{f}^{\phantom\dagger}_{\ve{i},s'}
\end{eqnarray}
\\
with the local constraint $\sum_{s}\hat{f}^{\dagger}_{\ve{i},s}  \hat{f}^{\phantom\dagger}_{\ve{i},s}=1$.
$s$- corresponds to a spin index and  $\ve{\sigma}$ corresponds to the vector of Pauli spin-1/2 matrices.

\begin{figure}[t]
\centering
\centerline{\includegraphics[width=0.35\textwidth]{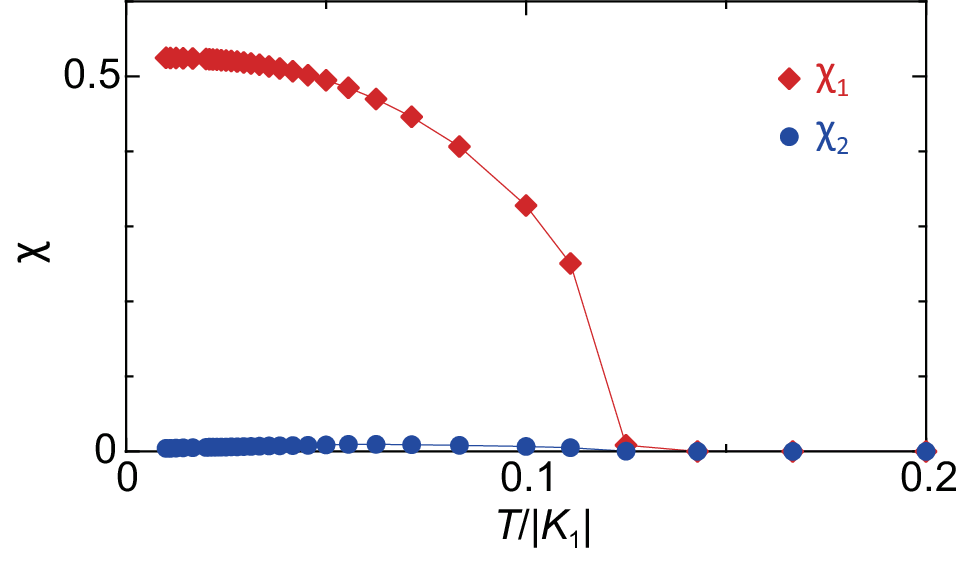}}
\caption{\label{fig:MFKitaev} 
Abrikosov fermionic mean-field results in the Kitaev model.
Here, $\chi_1=  \langle   \hat{D}^{\dagger}_{\ve{i},\ve{i}+\ve{\delta}_\gamma}  \rangle$ and $i \chi_2 =   \langle \hat{D}^{\gamma \dagger}_{\ve{i},\ve{i}+\ve{\delta}_\gamma}  \rangle$. 
}
\label{fig:unisus} 
\end{figure}

Using a  Fierz transformation an  exact  rewriting of    the  Kitaev  model  reads:   
\begin{eqnarray}
\hat{H}_{\text{Kitaev}}= &&-\frac{K_1}{8}\sum_{\ve{i}  \in A ,\gamma} \left(  \hat{D}_{\ve{i},\ve{i}+\delta_{\gamma}}^{\phantom\dagger}\hat{D}_{\ve{i},\ve{i}+\ve{\delta}_\gamma}^{\dagger}  +  \hat{D}_{\ve{i},\ve{i}+\delta_{\gamma}}^{\dagger}\hat{D}_{\ve{i},\ve{i}+\ve{\delta}_\gamma}^{\phantom\dagger} \right)  \nonumber\\ 
&&-\frac{K_1}{8}\sum_{\ve{i}  \in A ,\gamma} \left(  \hat{D}_{\ve{i},\ve{i}+\ve{\delta}_\gamma}^{\gamma}\hat{D}_{\ve{i},\ve{i}+\ve{\delta}_\gamma}^{\gamma \dagger}  +  
\hat{D}_{\ve{i}, \ve{i}+\ve{\delta}_\gamma }^{\gamma \dagger}\hat{D}_{\ve{i}, \ve{i}+\ve{\delta}_\gamma}^{\gamma} \right) \nonumber\\ 
\end{eqnarray}	
with
\begin{eqnarray}
\hat{D}_{\ve{i},\ve{i}+\ve{\delta}_\gamma} = \sum_{s}\hat{f}^{\dagger}_{\ve{i},s}  \hat{f}^{\phantom\dagger}_{\ve{i}+\ve{\delta}_\gamma,s}
\end{eqnarray}	
and 
\begin{eqnarray}
\hat{D}_{\ve{i},\ve{i}+\ve{\delta}_\gamma}^{\gamma} =\sum_{s,s'}\hat{f}^{\dagger}_{\ve{i},s} \sigma^{\gamma}_{s,s' }   \hat{f}^{\phantom\dagger}_{\ve{i}+\ve{\delta}_\gamma,s'}. 
\end{eqnarray}	
$ \hat{D}_{\ve{i},\ve{j}}$   accounts  for  a  spin-independent  hopping  between  sites  $\ve{i},\ve{j}$   such  that 
$  \hat{D}_{\ve{i},\ve{j}}  \hat{D}^{\dagger}_{\ve{i},\ve{j}}$   corresponds  to an  spin independent or  SU(2) invariant  exchange  process. 
$ \hat{D}^{\gamma}_{\ve{i},\ve{j}}$   accounts  for    spin-flip  processes  and  encodes  the  magnetic  anisotropy  of the  Kitaev  model.   
   A  mean  field  Ansatz   that  does  not  break  any   symmetries  of  the  original  Hamiltonian   and    leads  to 
a  mean  field  description    of  a  spin liquid  reads:  
$\chi_1 =  \langle   \hat{D}^{\dagger}_{\ve{i},\ve{i}+\ve{\delta}_\gamma}  \rangle$,  and  $i \chi_2 =   \langle \hat{D}^{\gamma \dagger}_{\ve{i},\ve{i}+\ve{\delta}_1}  \rangle $,   where both $\chi_1$  and    $\chi_2$   are  real.   
 The   result  of  the  mean-field   self-consistent  equations are  shown in 
Fig.~\ref{fig:MFKitaev}.    At  high   temperatures,   both  mean-field   order  parameters    vanish    accounting  for  the   high  temperature  independent
local  moment regime.   At  low   temperatures, $\chi_1 $  and  $ \chi_2 $   develop    finite   expectation values,  but with 
$\chi_1  \gg \chi_2 $.   

\end{document}